\documentstyle[aps,12pt]{revtex}
\newcommand{\bA}{\mbox{\boldmath $A$\unboldmath}}
\newcommand{\bn}{\mbox{\boldmath $n$\unboldmath}}
\newcommand{\bl}{\mbox{\boldmath $l$\unboldmath}}
\newcommand{\bm}{\mbox{\boldmath $m$\unboldmath}}
\newcommand{\cD}{\cal D}
\newcommand{\cL}{\cal L}
\newcommand{\cLd}{{\cal L}^{\dagger}}
\newcommand{\tkr}{2\kappa (r-r_H)}
\newcommand{\sta}{\sin\theta}
\newcommand{\cta}{\cos\theta}
\newcommand{\sda}{\sin^2\theta}

\newcommand{\coa}{\cot\theta}
\newcommand{\sqd}{\sqrt{2}}

\newcommand{\pr}{\frac{\partial}{\partial r}}
\newcommand{\pv}{\frac{\partial}{\partial v}} 
\newcommand{\pta}{\frac{\partial}{\partial \theta}}
\newcommand{\pvi}{\frac{\partial}{\partial \varphi}}
\newcommand{\pdr}{\frac{\partial^2}{\partial r^2}}
\newcommand{\pdta}{\frac{\partial^2}{\partial \theta^2}}
\newcommand{\pdvi}{\frac{\partial^2}{\partial \varphi^2}}
\newcommand{\pdvr}{\frac{\partial^2}{\partial v \partial r}}

\newcommand{\spr}{\frac{\partial}{\partial r_*}}
\newcommand{\spv}{\frac{\partial}{\partial v_*}}
\newcommand{\spdr}{\frac{\partial^2}{\partial r_*^2}}
\newcommand{\spdvr}{\frac{\partial^2}{\partial r_* \partial v_*}}

\begin{document}

\pagestyle{myheadings}
\markboth{}{Wu and Cai}
\draft

\title{\bf No New Quantum Thermal Effect of Dirac Particles \\
in a Charged Vaidya - de Sitter Black Hole}
\author{S. Q. Wu\thanks{E-mail: sqwu@iopp.ccnu.edu.cn} and 
X. Cai\thanks{E-mail: xcai@ccnu.edu.cn}}
\address{Institute of Particle Physics, Hua-Zhong 
Normal University, Wuhan 430079, China}
\date{\today}
\maketitle

\vskip 1cm
\begin{center}{\bf ABSTRACT}\end{center}
\begin{abstract}
\widetext
It is shown that Hawking radiation of Dirac particles does not exist for 
$P_1, Q_2$ components but for $P_2, Q_1$ components in a charged Vaidya - de 
Sitter black hole. Both the location and the temperature of the event horizon 
change with time. The thermal radiation spectrum of Dirac particles is the 
same as that of Klein-Gordon particles. Our result demonstrates that there 
is no new quantum effect in the thermal radiation of Dirac particles in any 
spherically symmetry black holes.

\end{abstract}

\newpage
\baselineskip 20pt

\section{Introduction}

In the last decades, the Hawking radiation \cite{Hawk} of Dirac particles had 
been largely investigated in some spherically symmetric and non-static black 
holes \cite{Zhetc}. However, most of these studies concentrated on the spin 
state $p=1/2$ of the four-component Dirac spinors. Recently, the Hawking 
radiation of Dirac particles of spin state $p= -1/2$ attracted a little more 
attention \cite{LZ}. In a series of papers, Li et al. \cite{LZ,LLM} claimed 
that they had discovered a kind of new quantum effect for the 
Vaidya-Bonner-de Sitter black hole. On the base of the generalized 
Teukolsky-type master equation \cite{Teuk} for fields of spin ($s = 0, 1/2, 
1$ and $2$ for the scalar, Dirac, electromagnetic and gravitational field, 
respectively) in the Vaidya-type space-times \cite{LZ,LLM}, they showed that 
this effect is depicted by
\begin{eqnarray}
\omega_0 &=& \frac{2(1-s-p)p\mu_0\lambda\dot{r}_H}{\lambda^2+(2p\mu_0r_H)^2}
\, , ~~(p = \pm s) \nonumber\\
&=& \left\{
\begin{array}{ll}
&\frac{-(\ell+1/2)\mu_0\dot{r}_H}{(\ell+1/2)^2+(\mu_0r_H)^2} \, ,
~~(\mu_0 \not= 0, s = -p = 1/2) \, , \\
&0 \, , ~~(\rm otherwise) \, .
\end{array}\right.
\end{eqnarray}
where $\mu_0$ is the mass of fields with spin-$s$.

Among their master equations, as the mass of Klein-Gordon particle and that
of Dirac particle are nonzero, so their argument sounds only for the massive 
spin-$1/2$ particles. They found that the massive Dirac field of spin state 
$p = 1/2$ differs greatly from that of spin state $p = -1/2$ in radiative 
mechanism, and suggested that it originate from the variance of Dirac vacuum 
near the event horizon in the non-static space-times caused by spin state. 
Further, they conjectured that this effect originates from the quantum 
ergosphere \cite{York}, that is, the quantum ergosphere can influence 
the radiative mechanism of a black hole.

In this paper, we re-investigate the Hawking effect of Dirac particles in the 
Vaidya-type black hole by means of the generalized tortoise transformation 
(GTCT) method. We consider simultaneously the limiting forms of the first 
order form and the second order form of Dirac equation near the event horizon 
because the Dirac spinors should satisfy both of them. From the former, we 
can obtain the event horizon equation, while from the latter, we can derive 
the Hawking temperature and the thermal radiation spectrum of electrons.
Our results are in accord with others. With our new method, we can prove 
rigorously that the Hawking radiation takes place only for $P_2, Q_1$ but
not for $P_1, Q_2$ components of Dirac spinors. The origin of this asymmetry 
of the Hawking radiation of different spinorial component maybe stem from the 
asymmetry of space-time in the advanced Eddington-Finkelstein coordinate system. 
As a by-product, we show that there could not have new quantum effect in the 
Hawking radiation of Dirac particles in any spherically symmetric black hole 
whether it is static or non-static. Our conclusion is contrary completely to 
that of Li's \cite{LZ,LLM} who argued that the radiative mechanism of massive 
spin fields depends on the spin state. 

The paper is organized as follows: In section 2, we write out the spinor form 
of Dirac equation in the Vaidya-type black hole, then, we obtain the event
horizon equation in Sec. 3. The Hawking temperature and the thermal radiation 
spectrum are derived in Sec. 4 and 5, respectively. Finally we present some 
discussions.

\section{Dirac equation}

The metric of a charged Vaidya - de Sitter black hole with the cosmological 
constant $\Lambda$ is given in the advanced Eddington-Finkelstein coordinate 
system by 
\begin{equation}
ds^2 = 2dv(G dv -dr) -r^2(d\theta^2 +\sin^2\theta d\varphi^2) \, , 
\end{equation}
and the electro-magnetic one-form is
\begin{equation}
\bA=\frac{Q}{r}dv \, 
\end{equation}
where $2G = 1 -\frac{2M}{r} +\frac{Q^2}{r^2} -\frac{\Lambda}{3}r^4$, in which 
both mass $M(v)$ and electric charge $Q(v)$ of the hole are functions of the 
advanced time $v$.

We choose such a complex null-tetrad $\{\bl, \bn, \bm, \overline{\bm}\}$  
that satisfies the orthogonal conditions $\bl \cdot \bn = -\bm \cdot 
\overline{\bm} = 1$. Thus the covariant one-forms can be written as
\begin{equation}
\begin{array}{ll}
\bl = dv \, , ~~
&\bm = \frac{-r}{\sqd}\left(d\theta +i\sta d\varphi\right) \, , \\
\bn = G dv -dr \, , ~~ 
&\overline{\bm} = \frac{-r}{\sqd}\left(d\theta -i\sta d\varphi\right) \, . 
\end{array}
\end{equation}
and their corresponding directional derivatives are
\begin{equation}
\begin{array}{ll}
D = -\pr \, ,  ~~
&\delta = \frac{1}{\sqd r}\left(\pta +\frac{i}{\sta}\pvi\right) \, , \\ 
\Delta = \pv +G\pr \, , ~~
&\overline{\delta} = \frac{1}{\sqd r}\left(\pta -\frac{i}{\sta}\pvi\right) \, .  
\end{array}
\end{equation}
 
The non-vanishing Newman-Penrose complex coefficients \cite{NP} in the above 
null-tetrad are easily obtained as follows 
\begin{equation}
\mu = \frac{G}{r} \, , ~~\gamma = -\frac{G_{,r}}{2} = -\frac{dG}{2dr} \, ,
~~\beta = -\alpha = \frac{\coa}{2\sqd r}  \, .  
\end{equation}

Inserting for the following relations among the Newman-Penrose  
spin-coefficients 
\begin{equation}
\begin{array}{ll}
\epsilon -\rho = -\frac{1}{r} \, , ~~
&\tilde{\pi} -\alpha = \frac{\coa}{2\sqd r} \, , \\
\mu -\gamma = \frac{G}{r} +\frac{G_{,r}}{2} \, , ~~
&\beta -\tau = \frac{\coa}{2\sqd r} \, ,
\end{array}
\end{equation}
and the electro-magnetic potential
\begin{equation}
\bA \cdot \bl = 0 \, ,~~\bA \cdot \bn = Q/r \, ,
~~\bA \cdot \bm = -\bA \cdot \overline{\bm} = 0 \, ,
\end{equation}
into the spinor form of the coupled Chandrasekhar-Dirac equation \cite{CP}, 
which describes the dynamic behavior of spin-$1/2$ particles, namely
\begin{equation}
\begin{array}{ll}
&(D +\epsilon -\rho +iq\bA\cdot\bl)F_1 +(\overline{\delta} +\tilde{\pi} 
-\alpha  +iq\bA\cdot\overline{\bm})F_2 =\frac{i\mu_0}{\sqd}G_1 \, , \\
&(\Delta +\mu -\gamma  +iq\bA\cdot\bn)F_2 +(\delta +\beta -\tau 
+iq\bA\cdot\bm)F_1 =\frac{i\mu_0}{\sqd}G_2 \, ,\\
&(D +\epsilon^* -\rho^* +iq\bA\cdot\bl)G_2 -(\delta +\tilde{\pi}^*
-\alpha^* +iq\bA\cdot\bm)G_1 =\frac{i\mu_0}{\sqd}F_2 \, , \\
&(\Delta +\mu^* -\gamma^* +iq\bA\cdot\bn)G_1 -(\overline{\delta}
+\beta^* -\tau^* +iq\bA\cdot\overline{\bm})G_2 =\frac{i\mu_0}{\sqd}F_1 \, ,
\end{array}
\end{equation}
where $\mu_0$ and $q$ is the mass and charge of Dirac particles, one obtains 
\begin{equation}
\begin{array}{ll}
-\left(\pr + \frac{1}{r}\right)F_1 +\frac{1}{\sqd r} {\cL}_{1/2} F_2 
= \frac{i\mu_0}{\sqd} G_1 \, , 
&\frac{1}{2r^2} {\cD} F_2  +\frac{1}{\sqd r} {\cLd}_{1/2} F_1 
= \frac{i\mu_0}{\sqd} G_2 \, , \\
-\left(\pr + \frac{1}{r}\right)G_2 -\frac{1}{\sqd r} {\cLd}_{1/2} G_1 
= \frac{i\mu_0}{\sqd} F_2 \, , 
&\frac{1}{2r^2} {\cD} G_1 -\frac{1}{\sqd r} {\cL}_{1/2} G_2 
= \frac{i\mu_0}{\sqd} F_1 \, , \label{DCP}
\end{array}
\end{equation}
in which we have defined operators
$${\cD} = 2r^2\left(\pv +G\pr \right) +(r^2G)_{,r} +2iqQr  \, ,$$
$${\cL}_{1/2} = \pta +\frac{1}{2}\coa -\frac{i}{\sta}\pvi  \, , 
~~{\cLd}_{1/2} = \pta +\frac{1}{2}\coa +\frac{i}{\sta}\pvi  \, .$$

By substituting
$$F_1 = \frac{1}{\sqd r} P_1 \, , ~~~F_2 = P_2 \, , ~~~G_1 = Q_1 \, , 
~~~G_2 = \frac{1}{\sqd r} Q_2 \, , $$
into Eq. (\ref{DCP}), they have the form
\begin{equation}
\begin{array}{rr}
-\pr P_1 +{\cL}_{1/2} P_2 = i\mu_0 r Q_1 \, , ~~
&{\cD} P_2 +{\cLd}_{1/2} P_1 = i\mu_0 r Q_2 \, , \\
-\pr Q_2 -{\cLd}_{1/2} Q_1 = i\mu_0 r P_2 \, , ~~
&{\cD} Q_1 -{\cL}_{1/2} Q_2 = i\mu_0 r P_1 \, . \label{reDP}
\end{array}
\end{equation}

\section{Event Horizon}

An apparent fact is that the Chandrasekhar-Dirac equation (\ref{reDP}) could 
be satisfied by identifying $Q_1$, $Q_2$, $qQ$ with $P_2^*$, $-P_1^*$, $-qQ$, 
respectively. So one may deal with a pair of components $P_1$, $P_2$ only. 
As to the thermal radiation, one may concern about the behavior of Eq. 
(\ref{reDP}) near the horizon. Though Eq. (\ref{reDP}) can be decoupled in a 
spherically symmetric space-time such as Vaidya black hole, we do not separate 
it in advance into a radial part and an angular one. As the Vaidya-type 
space-time is spherically symmetric, let's introduce a generalized tortoise 
coordinate transformation \cite{ZD} as 
\begin{equation}
r_* = r +\frac{1}{2\kappa}\ln[r -r_H(v)] \, , ~~v_* = v -v_0 \, ,\label{trans}
\end{equation}
where $r_H = r_H(v)$ is the location of the event horizon, $\kappa$ is an 
adjustable parameter and is unchanged under tortoise transformation. The 
parameter $v_0$ is an arbitrary constant. From formula (\ref{trans}), we 
can deduce some useful relations for the derivatives as follows:
$$ \pr = \left[1 +\frac{1}{\tkr}\right]\spr \, ,
~~\pv = \spv -\frac{r_{H,v}}{\tkr}\spr \, . $$

Under the transformation (\ref{trans}), Eq. (\ref{reDP}) with respect to a
pair of components ($P_1,P_2$) can be reduced to the following limiting form 
near the event horizon 
\begin{equation}
\begin{array}{ll}
&\spr P_1 = 0 \, , \\ 
& 2r_H^2\left[G(r_H) -r_{H,v}\right]\spr P_2 = 0 \, ,  
\label{trDPP} 
\end{array}
\end{equation}
after being taken the $r \rightarrow r_H(v_0)$ and $v \rightarrow v_0$ limits. 
A similar form holds for $Q_1, Q_2$.

From Eq. (\ref{trDPP}), we know that $P_1$ is independent of $r_*$ and regular 
on the event horizon. If the derivative $\spr P_2$ in Eq. (\ref{trDPP}) does 
not be equal to zero, the existence condition of a non-trial solution for $P_2$ 
is then (as for $r_H \not= 0$)
\begin{equation}
2G(r_H) -2r_{H,v} = 0 \, . \label{loca}
\end{equation}
which determines the location of horizon. The event horizon equation 
(\ref{loca}) can be inferred from the null hypersurface condition, 
$g^{ij}\partial_i F\partial_j F = 0$, and $F(v,r) = 0$, namely $r = r(v)$. 
Because $r_H$ depends on time $v$, so the location of the event 
horizon and the shape of the black hole change with time. 

\section{Hawking Temperature}

To investigate the Hawking radiation of $P_1, P_2$ components of spin-$1/2$ 
particles, one need consider the behavior of their second order form of
Dirac equation near the event horizon. A direct calculation gives the 
second order equation for ($P_1, P_2$)
\begin{eqnarray}
&&2r^2\pdvr P_1 +2r^2G\pdr P_1 +{\cL}_{1/2}{\cLd}_{1/2} P_1 -\mu_0^2 r^2 P_1 
\nonumber\\
&&~~+\left(2iqQr +r^2G_{,r} +2rG\right)\pr P_1  
+2ir^2G\mu_0 Q_1 = 0 \, , \label{socd+} \\
&&2r^2\pdvr P_2 +2r^2G\pdr P_2 +{\cLd}_{1/2}{\cL}_{1/2} P_2 
+\left(2iqQr +3r^2G_{,r} +6rG\right)\pr P_2  \nonumber \\
&&~~+4r\pv P_2 +(r^2G_{,rr} +4rG_{,r} +2G +2iqQ -\mu_0^2 r^2) P_2 
-i\mu_0 Q_2 = 0 \, . \label{socd-}
\end{eqnarray}
where 
\begin{eqnarray*}
{\cLd}_{1/2}{\cL}_{1/2} &=& \pdta +\coa \pta +\frac{1}{\sda}\pdvi +\frac{i\cta}{\sda}\pvi 
 -\frac{1}{4\sda} -\frac{1}{4} \, , \\
{\cL}_{1/2}{\cLd}_{1/2} &=& \pdta +\coa \pta +\frac{1}{\sda}\pdvi -\frac{i\cta}{\sda}\pvi 
-\frac{1}{4\sda} -\frac{1}{4}\, . 
\end{eqnarray*}
The angular parts of $P_1, P_2$ are spinorial spherical harmonics
${_{\pm 1/2}}Y_{\ell m}(\theta, \varphi)$ \cite{GMNRS}. One easily show
that 
$${\cL}_{1/2}{\cLd}_{1/2} P_1 = -(\ell +1/2)^2 P_1 \, , ~~~~
{\cLd}_{1/2}{\cL}_{1/2} P_2 = -(\ell +1/2)^2 P_2 \, .$$

Given the GTCT in Eq. (\ref{trans}) and after some calculations, the limiting 
form of Eqs. (\ref{socd+},\ref{socd-}), when $r$ approaches $r_H(v_0, \theta_0)$ 
and $v$ goes to $v_0$, reads
\begin{equation}
\left\{\frac{A}{2\kappa} +r_H^2[4G(r_H) -2r_{H,v}]\right\} \spdr P_1 
+2r_H^2 \spdvr P_1 = 0 \, , \label{wone+}
\end{equation}
and
\begin{eqnarray}
&&\left\{\frac{A}{2\kappa} +r_H^2[4G(r_H) -2r_{H,v}] \right\} \spdr P_2 
+2r_H^2 \spdvr P_2 \nonumber \\
&&~~~~ +\left\{-A +3r_H^2G_{,r}(r_H) +2iqQr_H +r_H[6G(r_H) -4r_{H,v}] 
\right\}\spr P_2 = 0 \, . \label{wone-} 
\end{eqnarray}
In deriving Eq. (\ref{wone+}) we have used the relation $\spr P_1 = 0$.

With the aid of the event horizon equation (\ref{loca}), namely 
$$2G(r_H) = 2r_{H,v} \, ,$$ 
\noindent
we know that the coefficient $A$ is an infinite limit of $0 \over 0$ type. 
By use of the L' H\^{o}spital rule, we get the following result
\begin{equation}
A = \lim_{r \rightarrow r_H(v_0)}\frac{2r^2(G -r_{H,v})}{r -r_H} 
= 2r_H^2G_{,r}(r_H) \, .
\end{equation}

Now let us select the adjustable parameter $\kappa$ in Eqs. (\ref{wone+},
\ref{wone-}) such that
\begin{equation}
r_H^2 \equiv \frac{A}{2\kappa} +2r_H^2[2G(r_H) -r_{H,v}] 
= \frac{r_H^2G_{,r}(r_H)}{\kappa} +2r_H^2G(r_H) \, ,
\end{equation}
which gives the temperature of the horizon 
\begin{equation}
\kappa =\frac{G_{,r}(r_H)}{1-2G(r_H)}= \frac{G_{,r}(r_H)}{1-2r_{H,v}} \, . 
\label{temp}
\end{equation}
With such a parameter adjustment and using the event horizon equation
(\ref{loca}), we can reduce Eqs. (\ref{wone+},\ref{wone-}) to
\begin{equation}
\spdr P_1 +2\spdvr P_1 = 0 \, ,  \label{wtwo+}
\end{equation}
and
\begin{equation}
\begin{array}{ll}
\spdr P_2 +2\spdvr P_2 +\left[G_{,r}(r_H) +2\frac{iqQ +G(r_H)}{r_H} 
\right]\spr P_2 = 0 \, . \label{wtwo-} 
\end{array}
\end{equation}
Eqs. (\ref{wtwo+},\ref{wtwo-}) are standard wave equations near the horizon. 
As the angular parts of $P_1, P_2$ have no relation to Hawking radiation, so 
we can omit them in the following section. The radial parts $R_1, R_2$ satisfy 
the same equations as that of $P_1 ,P_2$ 
\begin{eqnarray}
&&\spdr R_1 +2\spdvr R_1  = 0 \, , ~~~~\spr R_1 = 0 \, ,\label{wave+} \\
&&\spdr R_2 +2\spdvr R_2 +2\left(C +i\omega_0\right) \spr R_2 = 0 \, ,   
\label{wave-}
\end{eqnarray}
where $\omega_0, C$ will be regarded as finite real constants,
$$\omega_0 = \frac{qQ}{r_H} \, , 
~~~~C =\frac{1}{2} G_{,r}(r_H) +\frac{r_{H,v}}{r_H} \, .$$ 

\section{Thermal Radiation Spectrum}

From Eqs. (\ref{wave+}), we know that $R_1$ is a constant on the event 
horizon. The solution $R_1 = R_{10}e^{-i\omega v_*}$ means that Hawking 
radiation does not exist for $P_1, Q_2$. 

Now separating variables to Eq. (\ref{wave-}) as follows 
$$R_2 = R_2(r_*)e^{-i\omega v_*}$$ 
one gets
\begin{equation}
 R_2^{\prime\prime} = 2[i(\omega -\omega_0) -C] R_2^{\prime} \, , 
\end{equation}
The solution is
\begin{equation}
 R_2 \sim e^{2[i(\omega -\omega_0) -C]r_*} \, ; R_{20} \, . 
\end{equation}

The ingoing wave and the outgoing wave to Eq. (\ref{wave-}) are 
\begin{equation}
\begin{array}{ll}
&R_2^{\rm in} = e^{-i\omega v_*} \, ,\\
&R_2^{\rm out} = e^{-i\omega v_*} 
e^{2[i(\omega -\omega_0) -C]r_*} \, ,~~~~~~~ (r > r_H) \, . 
\end{array}
\end{equation}

Near the event horizon, we have 
$$r_* \sim \frac{1}{2\kappa}\ln (r - r_H) \, .$$
Clearly, the outgoing wave $R_2^{\rm out}(r > r_H)$ is not analytic at 
the event horizon $r = r_H$, but can be analytically extended from the 
outside of the hole into the inside of the hole through the lower complex 
$r$-plane
$$ (r -r_H) \rightarrow (r_H -r)e^{-i\pi}$$
to
\begin{equation}
\tilde{R}_2^{\rm out} = e^{-i\omega v_* }e^{2[i(\omega -\omega_0) -C]r_*}
e^{i\pi C/\kappa}e^{\pi(\omega -\omega_0)/\kappa} \, ,~~~~~~(r < r_H) \, . 
\end{equation}

So the relative scattering probability of the outgoing wave at 
the horizon is easily obtained
\begin{equation}
\left|\frac{R_2^{\rm out}}{\tilde{R}_2^{\rm out}}\right|^2
= e^{-2\pi(\omega -\omega_0)/\kappa} \, . 
\end{equation}

According to the method of Damour-Ruffini-Sannan's \cite{DRS}, the thermal 
radiation Fermionic spectrum of Dirac particles from the event horizon of the 
hole is given by
\begin{equation} 
\langle {\cal N}_{\omega} \rangle 
= \frac{1}{e^{(\omega -\omega_0)/T_H } + 1} \, , \label{sptr}
\end{equation} 
with the Hawking temperature being 
\begin{equation}
T_H = \frac{\kappa}{2\pi} 
= \frac{1}{4\pi r_H} \cdot \frac{M r_H -Q^2  
-\Lambda r_H^4/3 }{M r_H -Q^2/2 -\Lambda r_H^4/6 } \, .
\end{equation}
It follows that the temperature depends on the time, because it is determined 
by the surface gravity $\kappa$, a function of $v$. The temperature coincides 
with that derived from the investigating of the thermal radiation of 
Klein-Gordon particles \cite{LZ}.

\section{Conclusions}

In this paper, we have studied the Hawking radiation of Dirac particles in 
a black hole whose mass and electric charge change with time. We have dealt 
with the asymptotic behavior of Dirac equation near the event horizon, not 
only its first order form but also its second order form. We find that not 
all component of Dirac spinors but for $P_2, Q_1$ displays the property of 
thermal radiation. The asymmetry of Hawking radiation with respect to the 
four-component Dirac spinors maybe originate from the asymmetry of space-times 
in the advanced Eddington-Finkelstein coordinate. 

Equations (\ref{loca}) and (\ref{temp}) give the location and the temperature 
of event horizon, which depend on the advanced time $v$. They are just the 
same as that obtained in the discussing on thermal radiation of Klein-Gordon 
particles in the same space-time. Eq. (\ref{sptr}) shows the thermal radiation 
spectrum of Dirac particles in a charged Vaidya black hole with a cosmological 
constant $\Lambda$. These results are consistent with others.

From the thermal spectrum (\ref{sptr}) of Dirac particles, we know that 
there is no other interaction energy except the Coulomb energy $\omega_0$ 
in a Vaidya-type space-time. This inferres that there is no new quantum 
effect called by Li \cite{LZ,LLM} in any spherically symmetric black hole 
whether it is static or non-static. 

The discussion in this paper is easily extended to the case of the Hawking 
radiation of photon in the Vaidya-Bonner-de Sitter spacetime. Taking into
account of the restrict of the asymptotic behavior of the first order coupled 
Maxwell equations near the event horizon (namely, $\spr\phi_1 = \spr\phi_2 =0$), 
one can show that only the complex scalar $\phi_0$ takes part in the Hawking 
radiation and Li's conclusion does not hold in the case of particles with
spin-$1$. The results derived from the Hawking effect of photon are consistent 
with the present paper. Details will be published elsewhere.

\vskip 0.5cm
\noindent 
{\bf Acknowledgment}

S.Q. Wu is indebted to Dr. Jeff Zhao at Motomola Company for his longterm 
helps. This work is supported in part by the NSFC in China.


\end{document}